\documentclass[preprint,showpacs,preprintnumbers,amsmath,amssymb, showkeys]{revtex4}
\usepackage{array}
\usepackage{booktabs}
\usepackage{tabu}
\usepackage{dcolumn}
\usepackage{amsmath}
\usepackage{amsfonts}
\usepackage{amssymb}
\usepackage{graphicx}
\usepackage{subfigure}
\usepackage{graphicx}
\usepackage{dcolumn}
\usepackage{bm}

\begin{document}

\title{Hamilton-Jacobi Formalism for Tachyon Inflation}
\author{A. Aghamohammadi$^1$}
 \email{a.aqamohamadi@gmail.com}
  \author{A. Mohammadi$^2$}
   \email{abolhassanm@gmail.com}
    \author{T. Golanbari$^{3}$}
     \email{t.golanbari@uok.ac.ir}
      \author{Kh. Saaidi$^3$}
       \email{ksaaidi@uok.ac.ir}
\affiliation{
$^1$Sanandaj Branch, Islamic Azad University, Sanandaj, Iran\\
$^2$Young Researchers and Elites Club, Sanandaj Branch, Islamic Azad University, Sanandaj, Iran\\
$^3$Department of Physics, Faculty of Science, University of Kurdistan,  Sanandaj, Iran}
\date{\today}

\def\be{\begin{equation}}
  \def\ee{\end{equation}}
\def\bea{\begin{eqnarray}}
\def\eea{\end{eqnarray}}
\def\f{\frac}
\def\n{\nonumber}
\def\l{\label}
\def\p{\phi}
\def\o{\over}
\def\R{\rho}
\def\pa{\partial}
\def\om{\omega}
\def\na{\nabla}
\def\P{\Phi}

\begin{abstract}
Tachyon inflation is reconsidered by using the recent observational data obtained from Planck-2013 and BICEP2. The Hamilton-Jacobi formalism is picked out as a desirable approach in this work, which allows one to easily obtain the main parameters of the model. The Hubble parameter is supposed as a power-law and exponential function of the scalar field, and each case is considered separately. The constraints on the model, which come from observational data, are explained during the work. The results show a suitable value for tensor spectral index and an appropriate form of the potential.
\end{abstract}
\maketitle

\section{Introduction}
Inflationary cosmology has become the unrivaled paradigm for explaining the early evolution of the Universe. Over past decades, this scenario has been supported by different observational data \cite{1}. It could perfectly describe the problem of hot big bang model, and also provides a mechanism for producing density perturbations, which are absolutely necessary for large scale structure of the Universe \cite{1,2}. It also predicts a gravitational wave which has just been confirmed \cite{3}. Standard inflationary cosmology is described by a single scalar field which slowly rolls down its potential. This scenario is known as chaotic inflation proposed by Linde in 1983 \cite{4}.  So far, different inflationary scenarios have been introduced; however, chaotic inflation has become the most favored one because of its simplicity. Various potentials which give a desirable inflation, have been considered \cite{5,6,7}. In addition to the slow-rolling approximation, there is another method for studying inflation known as the Hamilton-Jacobi approach \cite{8,9,10,11}. In this approach, the Hubble parameter is introduced as a function of the scalar field, instead of introducing a potential. By doing so, the whole parameters of the model could be derived easily.

M or String theory inspired models are under active consideration in cosmology at present. It has recently been suggested that the rolling tachyon condensate, in a class of string theories, may have interesting cosmological consequences \cite{12}. Sen \cite{13} has shown that the decay of D-branes produces a pressureless gas with finite energy density that resembles classical dust.
Rolling tachyon matter associated with unstable D-branes has an interesting equation of state which smoothly interpolates between  -1 and 0 \cite{12}. The tachyon field associated with unstable D-branes might be responsible for cosmological inflation at early epochs due to tachyon condensation near the top of the effective potential \cite{14,15} and could contribute to some new form of cosmological dark matter at late times \cite{13}. The cosmological aspect of tachyons has been studied by several authors \cite{14,15,16,17,18,19,20,21,22,23,24,25,26,27,28,29,30,31,32}. Besides the string theory aspect of this model, tachyon inflation has been considered by utilizing potentials which are related to the \textit{k}-inflation model \cite{33,34,35}.

In the present work, we consider tachyon inflation by using the Hamilton-Jacobi approach. The relevant parameters are obtained in terms of the Hubble parameter and its first derivative. It shall be shown that all of the main parameters of the model could easily be derived. The latest observational data are used to determine the free parameters of the model. the biggest problem in the study of scalar field models is unknown potential form. Based on this, we will obtain some potential forms which have the best agreement with observational data in the inflation epoch. \\
The paper is organized as follows: Sec.II is related to the general equation of the model. The Friedmann equations and conservation equation are expressed in this section. Then, in two subsections, the Hamilton-Jacobi and attractive behavior of the model are discussed. In Sec.III, for more specific results, two typical functions are introduced for the Hubble parameter in two separate subsection, as power-law and exponential function, and the observational constraints on the model are explained. The results of the work are summarized in Sec.IV.
\section{General Framework}
In Born-Infeld form, the effective action for tachyon could be described by \cite{36}
\begin{equation}\label{01}
\mathcal{S} = \int d^4x \sqrt{-g} \; \left( {m_p^2 \over 16\pi} R - V(\phi)\sqrt{1 + g^{\mu\nu} \partial_\mu\phi \partial_\nu\phi} \right),
\end{equation}
in which $R$ is the Ricci scalar, derived from metric $g_{\mu\nu}$, and $\phi$ is the tachyon scalar field with minimal coupling to gravity. The potential of the scalar field is denoted by $V(\phi)$. Also, $m_p$ is the Planck mass. In this work, we are going to consider the Universe evolution at one of the earliest times, namely, inflation. In this period of time, it is assumed that the scalar field dominates the Universe and derives inflation.\\
Taking a variation of action with respect to two independent variables $g_{\mu\nu}$ and $\phi$ leads to two main dynamical equations. Based on recent observational data, the Universe is isotropic, homogeneous, and spatially flat. The common metric to describing such a geometry is the Friedmann-Lemaitre-Robertson-Walker (FLRW) metric, which is read as
\begin{equation}\label{02}
ds^2 = -dt^2 + a^2(t) \left( dx^2 + dy^2 + dz^2 \right).
\end{equation}
The energy-momentum tensor for such a spatially flat FLRW universe is described by $T_{\mu}^{\nu}={\rm diag}(-\rho,p,p,p)$, where $\rho$ and $p$ are tachyon energy density and pressure, respectively and given by
\begin{equation}\label{03}
\rho = {V(\phi) \over \sqrt{1-\dot\phi^2}} , \qquad  p = -V(\phi) \sqrt{1 - \dot\phi^2}.
\end{equation}
The Friedmann equation keeps its usual form as
\begin{equation}\label{04}
H^2 = {8\pi \over 3m_p^2} \; \rho , \qquad \dot{H} = {-4\pi \over m_p^2} (\rho + p),
\end{equation}
and the acceleration could be derived easily from Eq.\;(\ref{04}) as
\begin{equation}\label{05}
{\ddot{a} \over a} = H^2 + \dot{H} = {8\pi\rho \over 3m_p^2} \; (1 - {3 \over 2} \dot\phi^2).
\end{equation}
The Universe stays in the accelerated expansion phase as long as $\dot\phi^2 < 2/3$ . This condition is different from its corresponding condition in standard inflation where $\dot\phi^2<V(\phi)$ because there is no mention of a potential \cite{38}. \\
The equation of motion of the tachyon scalar field comes from Eq.(\ref{01}) as
\begin{equation}\label{06}
\ddot\phi + \Big( 3H\dot\phi + {V'(\phi) \over V(\phi)} \Big) (1 - \dot\phi^2) = 0,
\end{equation}
which could be rewritten as a familiar form of the conservation equation $\dot\rho + 3H\rho(1+\omega)=0$. The tachyon inflation should begin with a very small value of $\dot\phi$, namely, $\dot\phi^2 \ll 1$, to get enough inflation. \\
There are two known approaches for studying the inflation scenario. The  first approach is "slow-rolling," which has been widely used. In this formalism, it is assumed that the potential of the scalar field dominated over the kinetic term, and the scalar field slowly rolls down to the minimum of its potential. The slow-rolling approximation comes to an almost flat potential during inflation, which leads to enough expansion to solve standard cosmology problems. Moreover, it is necessary to introduce a form of the potential to arrive at the final results of the model. The second approach is known as Hamilton-Jacobi. Instead of a potential, we need only to introduce a function of the scalar field for Hubble parameter $H:=H(\phi)$. Then, one can derive almost the whole of the parameters of the model. In the present work, the Hamilton-Jacobi approach is picked out for considering the Universe evolution in the inflationary epoch.\\

\subsection{Hamilton-Jacobi approach}
The Hubble parameter is supposed as a function of the scalar field, namely, $H:=H(\phi)$. Then, the time variable of $H$ could be rewritten as $\dot{H}=\dot\phi H'$, where the prime denotes derivative with respect to the scalar field. By using Eqs.\;(\ref{03}) and (\ref{04}) , the time derivative of the scalar field could be obtained in terms of scalar field as follows
\begin{equation}\label{07}
\dot\phi = -{2 \over 3} {H'(\phi) \over H^2(\phi)}.
\end{equation}
It is clear that, if $H'<0$ ($H'>0$), the scalar field increases (decreases) by passing time, namely, $\dot\phi>0$ ($\dot\phi<0$). \\
Using Eq.\;(\ref{07}) and Friedmann equations (\ref{04}), one can obtain
\begin{equation}\label{07a}
[H'(\phi)]^{2}-{9 \over 4}H^{4}(\phi)+({4 \pi \over m^{2}_{p}})^{2}V^{2}(\phi)=0.
\end{equation}
Equation (\ref{07a}) is the Hamilton-Jacobi equation. The potential of the model is easily obtained as a function of the scalar field from Eq. (\ref{07a}) as
\begin{equation}\label{08}
V(\phi) = {3m_p^2 \over 8\pi} H^2(\phi) \; \sqrt{1 - {4 \over 9} {H'^2(\phi) \over H^4(\phi)}}.
\end{equation}
As the tachyon field grows up, its potential tends to zero [$\phi \rightarrow \infty \quad $ and  $ \quad V(\phi) \rightarrow 0$];, however, the exact form of the potential is unknown yet \cite{39,40,41}. It was argued that the exponential potential could explain the qualitative dynamics of string theory of the tachyon \cite{36}. Later, it was suggested that a desirable cosmology could be constructed by a runaway potential with the tachyonic equation of state \cite{42}.\\
To avoid an imaginary potential, the term $4H'^2/9H^4$ must always be smaller than unity. Therefore, it comes to a condition as $4H'^2/9H^4 < 1$, which must be satisfied. \\
In comparison to ordinary single scalar field inflation, the slow-roll parameters are defined by \cite{37}
\begin{equation}\label{09}
\epsilon(\phi) \equiv {2 \over 3} \left( {H'(\phi) \over H^2(\phi)} \right)^2 , \qquad \eta(\phi) \equiv {1 \over 3} \left( H''(\phi) \over H^3(\phi) \right).
\end{equation}
It is assumed that during inflation the Universe undergoes a quasi-de Sitter expansion, in which the slow-roll parameters are much smaller than unity: $\epsilon, |\eta| \ll 1$ (known as the slow-roll approximation). Inflation ends when $\ddot{a}$ vanishes or, equivalently, when the slow-roll parameter $\epsilon$ arrives at unity. Hence, at the end of inflation, we have
\begin{equation}\label{10}
H^4 = {2 \over 3} \; H'^2;
\end{equation}
therefore, the scalar field at the end of inflation could easily be read.

Another important parameter is the scale factor of the Universe, which is denoted by $a(t)$. By using Eq.\;(\ref{07}) and relation $\dot{a}=\dot\phi a'$, the scale factor is expressed as
\begin{equation}\label{11}
a(\phi) = a_c \exp \left( -{3 \over 2}\; \int {H^3 \over H'} d\phi \right),
\end{equation}
where $a_c$ is the constant of integration. The number of \textit{e}-folds, which describe the amount of expansion, is described as
\begin{equation}\label{12}
N \equiv \int^{t_e}_{t_i} H dt = \int^{\phi_e}_{\phi_i} {H(\phi) \over \dot\phi} d\phi.
\end{equation}
in which the subscript "\textit{e}" and "\textit{i}," respectively, denote the end and beginning of inflation. It seems that, in the Hamilton-Jacobi approach, the main parameters of model could be derived more easily than the slow-rolling approach, with less assumptions.\\
During inflation, the scalar field and gravitons underwent quantum fluctuations. One of the most important advantages of inflation is that this theory not only can solve the problem of standard cosmology, it also is able to provide a mechanism to explain the observed anisotropy. The main fluctuation are known as scalar and tensor fluctuations, which are, respectively, known as curvature and gravitational perturbationa. These perturbation for tachyon inflation are derived by using a similar method as original scalar field inflation. The scalar perturbation for the tachyon scalar field gets a different equation because of its different coupling to geometry. However, the tensor perturbation part, which describes the propagation of gravitational waves, possesses the same equation as the original scalar field, since there it has no coupling to matter. Both of these equations and their answers are discussed in Ref. \cite{38} in more detail.
\subsection{Attractor behavior}
By using the Hamilton-Jacobi approach, it is easy to consider whether all possible trajectories (or solutions) converge to a common attractor solution. Doing so, we are going to follow Ref. \cite{43} and assume a homogeneous perturbation to a solution $H_0(\phi)$. If the homogenous perturbation part $\delta H$ becomes small by passing time, it could be said that the attractor condition is satisfied. \\
Substituting $H(\phi) = H_0(\phi) + \delta H(\phi)$ into Eq.\;(\ref{07a}) and linearizing, one arrives at
\begin{equation}\label{15}
H_0^3(\phi) \delta H(\phi) - {2 \over 9}\; H'_0(\phi) \delta H'(\phi) \simeq 0.
\end{equation}
The solution of the above differential equation is given by
\begin{equation}\label{16}
\delta H(\phi) = \delta H(\phi_i) \exp \left( {9 \over 2} \int^{\phi}_{\phi_i} {H_0^3(\phi) \over H'_0(\phi)} d\phi \right),
\end{equation}
where $\delta H(\phi_i)$ is the initial value of perturbation at $\phi=\phi_i$. Having $H(\phi)$, one could investigate the behavior of perturbation $\delta H(\phi)$. the Hamilton-Jacobi approach makes it easy to consider attractive behavior of solutions.\\
In the following section, a few typical examples are expressed for $H(\phi)$, and the consequences will be discussed in more detail. \\

\section{Typical Examples}
Up to now, the general forms of parameters have been derived, and some crude results were obtained. In this section, in order to get more specific results, we are going to introduce two functions for the Hubble parameter in terms of the scalar field, as a power-law function and an exponential function.\\

\subsection{Power-law function}
As the first case, we assume the Hubble parameter is a power-law function of scalar field $H(\phi)=\mathcal{H}_1 \phi^n$, where $\mathcal{H}_1$ and $n$ are constant. By using Eq.\;(\ref{07}), the scalar field could be expressed by
\begin{equation}\label{17}
\dot\phi = -{2 \over 3} {n \over \mathcal{H}_1 \phi^{n+1}}.
\end{equation}
Therefore, it clearly could be seen that, the scalar field has a decreasing (increasing) behavior for $n>0$ ($n<0$). Regardless of the $n$ sign, the general form of potential could be derived from  Eq.\;(\ref{08}):
\begin{equation}\label{18}
V(\phi) = {3m_p^2 \over 8\pi} \mathcal{H}_1^2 \phi^{2n} \sqrt{1 - {4 \over 9} {n^2 \over \mathcal{H}_1^2 \phi^{2(n+1)}}}.
\end{equation}
The condition of having a real potential leads to the following expression:
\begin{equation}\label{19}
\phi^{2(n+1)} > {4n^2 \over 9\mathcal{H}_1^2},
\end{equation}
which should be satisfied during the inflation. It easily could be checked that the potential condition is satisfied for all values of $n$ except for $0<n<-1$, where one encounters an imaginary potential. \\
Inflation ends when the acceleration $\ddot{a}$ vanishes. The scalar field at the end of inflation could be obtained from Eq.\;(\ref{10}) as
\begin{equation}\label{20}
\phi_e^{2(n+1)} = {2 \over 3} \; {n^2 \over \mathcal{H}^2_1}.
\end{equation}
To get the initial value of the scalar field, the number of \textit{e}-folds relation (\ref{12}) could be used. The integration is computed easily, and one arrives at
\begin{equation}\label{21}
\phi_i^{2(n+1)} = {2n^2 \over 3\mathcal{H}^2_1} \; \mathcal{N},
\end{equation}
where
\begin{equation}\nonumber
\mathcal{N} = \Big[ {2(n+1)N \over n} + 1 \Big].
\end{equation}
The slow-roll parameters (\ref{09}) for this case are
\begin{equation}\label{22}
\epsilon(\phi) = {2n^2 \over 3\mathcal{H}_1^2 \phi^{2(n+1)}}, \qquad \eta(\phi)= {n(n-1) \over 3\mathcal{H}_1^2 \phi^{2(n+1)}}.
\end{equation}
The scalar and tensor spectra indices, which directly come from the amplitude of scalar and tensor perturbations \cite{38, Per}, are easily derived as
\begin{eqnarray}
n_s - 1 & = & 4\eta(\phi_i) - 6\epsilon(\phi_i) = {-(4n+2) \over n\mathcal{N}} , \label{23}\\
 \ n_T \ & = &  -2 \epsilon(\phi_i) = {-2 \over \mathcal{N}},\label{24}
\end{eqnarray}
and the consistency relation could be written as $r=-8n_T$ (for more details, refer to Refs. \cite{38, Per}).
According to recent observational data obtained from WMAP9 + eCMB + BAO + $ H_{0} $, the Planck-2013 satellite and  Planck2013 + WP + highL + BAO, the scalar spectral index is about $n_s = 0.9608 \pm 0.0080 $, $ 0.9635 \pm 0.0094$, and $ 0.9608 \pm 0.0054 $, respectively. These observational results could be used to impose a constraint on $n$, and then one can obtaine the tensor spectral index, predicted by the model. In Table.\ref{Tab01}, the variable $n$ and tensor spectral index $n_T$ are derived for three different values of $n_s$ and number of \textit{e}-folds $N$. It is seen that, the parameter $n$ gets a negative value whose magnitude  decreases by increasing the number of \textit{e}-folds, and increases by a little enhancement in the scalar spectral index.  The tensor spectral index is predicted to be about $-0.02$, which is acceptable in comparison to observational data. In contrast with parameter $n$, the magnitude of the tensor spectral index increases by enhancement of $N$ and reduces by increasing $n_s$.

Another perturbation is tensor perturbation, which is also known as gravitational waves. The produced tensor
fluctuations induce a curled polarization in the CMB radiation and increase the overall amplitude of its anisotropies at
a large scale. The physics of the early Universe could be specified by fitting the analytical
results of the CMB and density spectral to corresponding observational data. At first, it was
thought that the possible effects of primordial gravitational waves are not important and
might be ignored. However, a few years ago, it was found out that the tensor
fluctuations have an important role and should be more attended for determining best-fit values of
the cosmological parameters from CMB and LSS spectra \cite{44,45,46}. The imprint of tensor fluctuation on the CMB brings this
idea to indirectly determine its contribution to power spectra by measuring CMB polarization \cite{45}. Such a
contribution could be expressed by the $r$ quantity, which is known as the tensor-to-scalar ratio and represents the
relative amplitude of the tensor to scalar fluctuation: $r={A_T^2 / A_s^2}$. Therefore, constraining $r$ is one of the main goals of
a modern CMB survey.
On the other hand, determining an exact value for $r$, which represents the existence of gravitational waves, is very difficult. Based on Planck satellite data, we were able only to put an upper bound for $r$ as $r<0.11$. A year later, BICEP2 achieved this goal and found out an exact value for this quantity as $r=0.20$, which is our latest observational data.

The amplitude of the tensor perturbation is given by $A_T^2 = r A_s^2$, where $A_s^2$ is the amplitude of the scalar perturbation. Based on Planck-2013 data, this quantity is about $\ln(10^{10}A_s^2)=3.098$ \cite{47}. The differential equation of the tensor perturbation for the tachyon scalar field is the same as the ordinary scalar field, because there is no coupling between the tensor perturbation and matter term. Then one could arrive at the same results for the amplitude of the tensor perturbation as $A_T^2=4H^2/25\pi m_p^2$ \cite{38, 48}. By using this relation, and the newest observational data about $A_s^2$ and $r$, the quantity $\mathcal{H}_1$ could be determined.
In Table \ref{Tab01}, the values of free parameters $\mathcal{H}_1$ are presented for different values of the number of \textit{e}-folds and spectral index, by using recent observational data. $\mathcal{H}_1$ increases by enhancement of the number of \textit{e}-folds, and, for each value of $N$, it decreases by increasing the scalar spectral index.

\begin{widetext}
\begin{figure}[ht]
\centering
\subfigure[Potential for $N=55$]{\includegraphics[width=5cm]{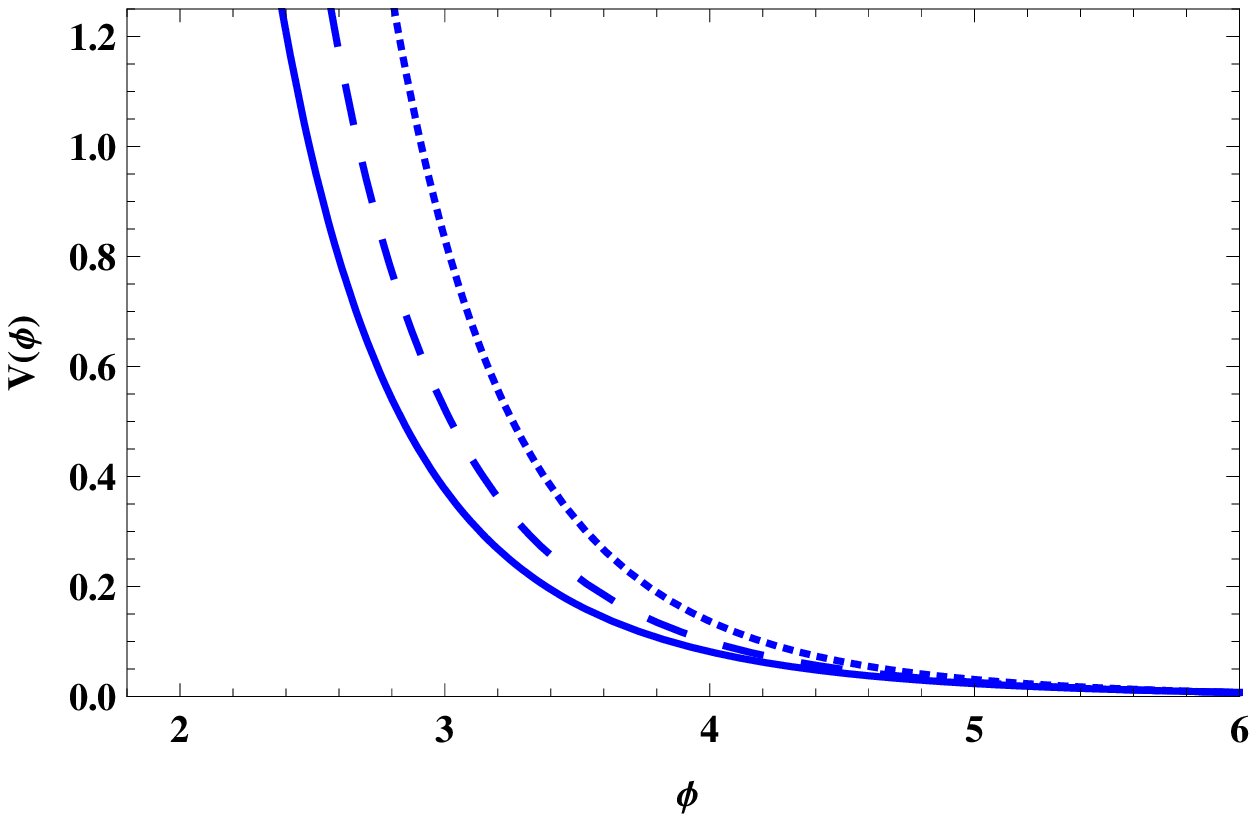}\label{FC01a} }
\hspace*{5mm}
\subfigure[Potential for $N=60$]{\includegraphics[width=5cm]{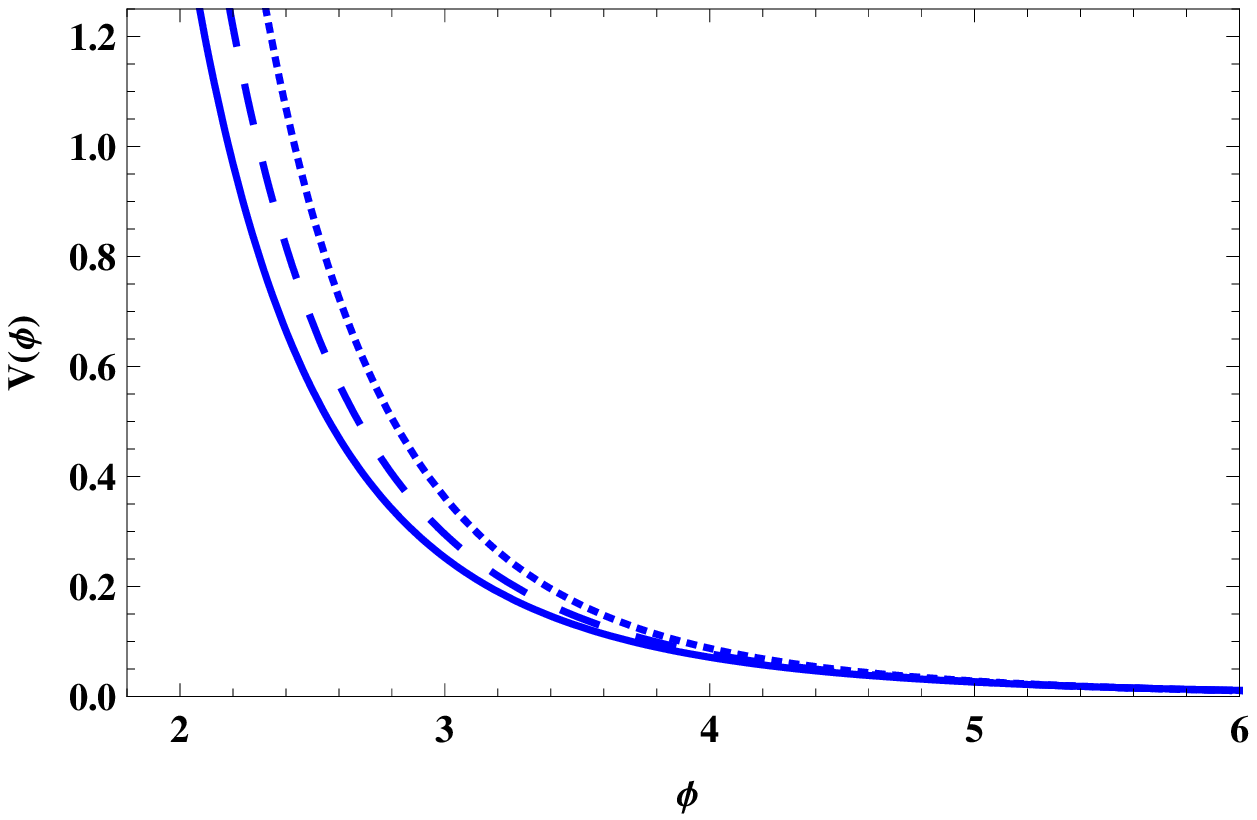}\label{FC01b} }
\hspace*{5mm}
\subfigure[Potential for $N=65$]{\includegraphics[width=5cm]{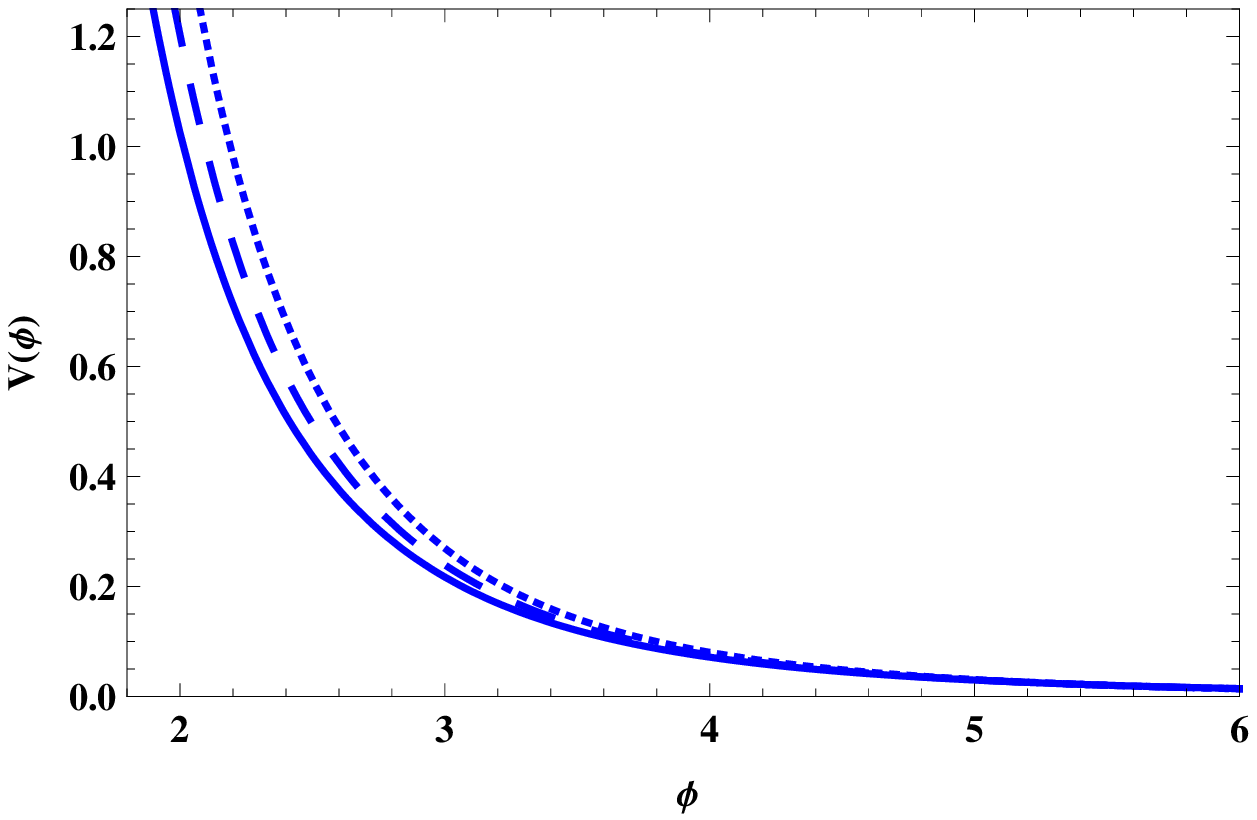}\label{FC01c} }
\caption{ The potential for different values of the number
of \textit{e}-folds and scalar spectral index as n$_s=0.9528$ (solid line), $0.9541$ (large-dashed line), and $0.9554$ (dotted line). The vertical axis indicates the potential ($\times 10^{-45}$), and the horizon axis denotes the scalar field ($\times 10^{-4}$).}\label{F01}
\end{figure}
\end{widetext}

The free parameters of the model have been determined from observational data and are expressed in Table \ref{Tab01}. These results could be used in potential equation (\ref{18}), to find out which kind of potential has been predicted by the model for this case. The potential is depicted for three different values of $N$ in Figs. \ref{FC01a}, \ref{FC01b}, and \ref{FC01c}. In each plot, we have three potential curves corresponding to three different values of $n_s$. A negative value of $n$ comes to a decreasing potential which approaches to zero by increasing the scalar field. It is clearly seen that the gap between these curves reduces by enhancement of $N$, and they come close to each other.
\subsubsection{Attractive behavior}
As a final step of the first typical example, the attractor behavior of the solution is considered. From Eq.\;(\ref{16}), the perturbation $\delta H(\phi)$ is obtained as
\begin{equation}\label{26}
\delta H(\phi) = \delta H(\phi_i) \exp \left( {9\mathcal{H}_1^2 \over 4n(n+1)} \Big[ \phi^{2(n+1)} - \phi_i^{2(n+1)} \Big] \right).
\end{equation}
For negative values of $n$ in Table \ref{Tab01}, the time derivative of scalar field is positive, which shows that the scalar field increases by passing time. Therefore, the exponential term on the right-hand side of Eq.(\ref{26}) decreases by passing time and tends to zero, then the perturbation part of the Hubble parameter vanishes, and the model has an attractive behavior. \\
\begin{widetext}
\begin{table}[ht]
  \centering
  \begin{tabular}{p{1.1cm}||p{2cm}p{2.5cm}p{2.5cm}p{2.5cm}p{2.5cm}}
    \toprule[1.5pt] \\[-2mm]
           & $n_s$ &  $0.9528$ \qquad  & $0.9541$ \qquad  & $0.9554$  \\[0.5mm]
      \midrule[1.5pt] \\[-1mm]
    $N=55$ \qquad   & \ $n$             \quad & $-2.57$ & $-2.78$  & $-3.05$ \\[2mm]
                    & \ $n_T$           \quad & $-0.029$ & $-0.027$ & $-0.026$ \\[2mm]
                    & \ $\mathcal{H}_1$ \quad & $1.97\times 10^{-5}$ & $4.25\times 10^{-6}$ & $5.88\times 10^{-7}$ \\[1.5mm]
    \hline \\[-3mm]
    \hline \\[-1mm]
    $N=60$ \qquad   & \ $n$             \quad & $-2.14$  & $-2.25$  & $-2.40$ \\[2mm]
                    & \ $n_T$           \quad & $-0.030$ & $-0.029$ & $-0.028$ \\[2mm]
                    & \ $\mathcal{H}_1$ \quad & $5.49\times 10^{-4}$ & $2.3\times 10^{-4}$ & $8.03\times 10^{-5}$\\[1.3mm]
    \hline \\[-3mm]
    \hline \\[-1mm]
    $N=65$ \qquad   & \ $n$             \quad & $-1.89$  & $-1.97$  & $-2.06$  \\[2mm]
                    & \ $n_T$           \quad & $-0.032$ & $-0.030$ & $-0.029$ \\[2mm]
                    & \ $\mathcal{H}_1$ \quad & $3.76\times 10^{-3}$ & $2.12\times 10^{-3}$ & $1.08\times 10^{-4}$ \\[1.5mm]
    \bottomrule[1.5pt]
  \end{tabular}
  \caption{Constraint on variable $n$, tensor spectral index $n_T$, and free parameter $\mathcal{H}_1$ for three different values of the number of \textit{e}-folds and scalar spectral index. The scalar spectral index is predicted by WMAP9 + eCMB + BAO + $H_{0}$ about $n_s = 0.9608 - 0.0080 = 0.9528$, Planck-2013 about $n_s = 0.9635 - 0.0094 = 0.9541$, and Planck2013 + WP + highL + BAO about $n_s = 0.9608 - 0.0054 = 0.9554$, and,  based on the latest observational results of Planck-2013, the tensor-to-scalar ratio is taken as $r=0.10$.}\label{Tab01}
\end{table}
\end{widetext}
\subsection{Exponential function}
In the second case, an exponential function of the scalar field is introduced for the Hubble parameter as $H(\phi) = \mathcal{H}_2 \exp(s\phi)$, where $s$ is constant and it could be positive or negative. The time derivative of the scalar field is read from Eq.\;(\ref{07}):
\begin{equation}\label{27}
\dot\phi = {-2s \over 3\mathcal{H}_2} \ \exp(-s\phi),
\end{equation}
so the scalar field increases (decreases) for negative (positive) values of $s$. The potential for this case is expressed as follows:
\begin{equation}\label{28}
V(\phi) = {3m_p^2 \over 8\pi} \mathcal{H}_2^2 \exp(2s\phi)\; \sqrt{1 - {4s^2 \over 9\mathcal{H}_2^2} \exp(-2s\phi)},
\end{equation}
which remains real during inflation. By the same method, at the end of inflation, by using Eq. (\ref{10}), the scalar field is derived as
\begin{equation}\label{29}
\exp(2s\phi_e) = {2s^2 \over 3\mathcal{H}_2^2},
\end{equation}
and, on the other hand, the initial scalar field is given in terms of the number of \textit{e}-folds by
\begin{equation}\label{30}
\exp(2s\phi_i) = {2s^2 \over 3\mathcal{H}_2^2} \ (1+2N).
\end{equation}
From Eqs.\;(\ref{09}), the slow-roll parameters are expressed by
\begin{equation}\label{31}
\epsilon(\phi) = {2s^2 \over 3\mathcal{H}_2^2 \exp(2s\phi)}, \qquad \eta(\phi)= {s^2 \over 3\mathcal{H}_2^2 \exp(2s\phi)}.
\end{equation}
Then scalar and tensor spectral indices related to this case are obtained by using the above relation and initial value of the scalar field as
\begin{eqnarray}
n_s - 1 & = & 4\eta(\phi_i) - 6\epsilon(\phi_i) = {-4 \over 1+2N} , \label{32}\\
 \ n_T \ & = &  -2 \epsilon(\phi_i) = {-2 \over 1+2N},\label{33}
\end{eqnarray}
which, in contrast to the previous case, depend only on the number of \textit{e}-folds parameter.
\begin{widetext}
\begin{table}[ht]
  \centering
  \begin{tabular}{lp{2.5cm}p{2.5cm}p{2.5cm}p{2.5cm}}
     \toprule[1.5pt] \\ [-2mm]
         $n_s$     & $0.9528$ \qquad  & $0.9611$ \qquad  & $0.9675$ \qquad \\[0.5mm]
              \midrule[1pt] \\[-2mm]
     $N$ \qquad\qquad &  $41.87$     \qquad  & $50.91$     \qquad &  $61.03$    \qquad \\[2mm]
     $n_T$ \qquad\qquad  & $-0.023$  \qquad  & $-0.019$    \qquad & $-0.016$    \qquad \\[2mm]
     $s / m_p$ \qquad\qquad  & $\pm 8.77\times 10^{-6}$ \qquad  & $\pm 7.96\times 10^{-5}$ \qquad & $\pm 7.28\times 10^{-5}$ \qquad \\[1.5mm]
     \bottomrule[1.5pt]
   \end{tabular}
  \caption{\footnotesize The constraint on the number of \textit{e}-folds $N$ and variable $s$ for different values of $n_s$. The scalar spectral index is predicted by WMAP9 + eCMB + BAO + $H_{0}$ about $n_s = 0.9608 - 0.0080 = 0.9528$, Planck2013 + WP + highL + BAO about $n_s = 0.9611$, and Planck-2013 about $n_s =0.9675$, and, based on the latest observational results of Planck-2013, the tensor-to-scalar ratio is taken as $r=0.10$.}\label{Tab03}
\end{table}
\begin{figure}[ht]
\centering
\subfigure[Potential for $\mathcal{H}_2^{2}=10^{30}$]{\includegraphics[width=5cm]{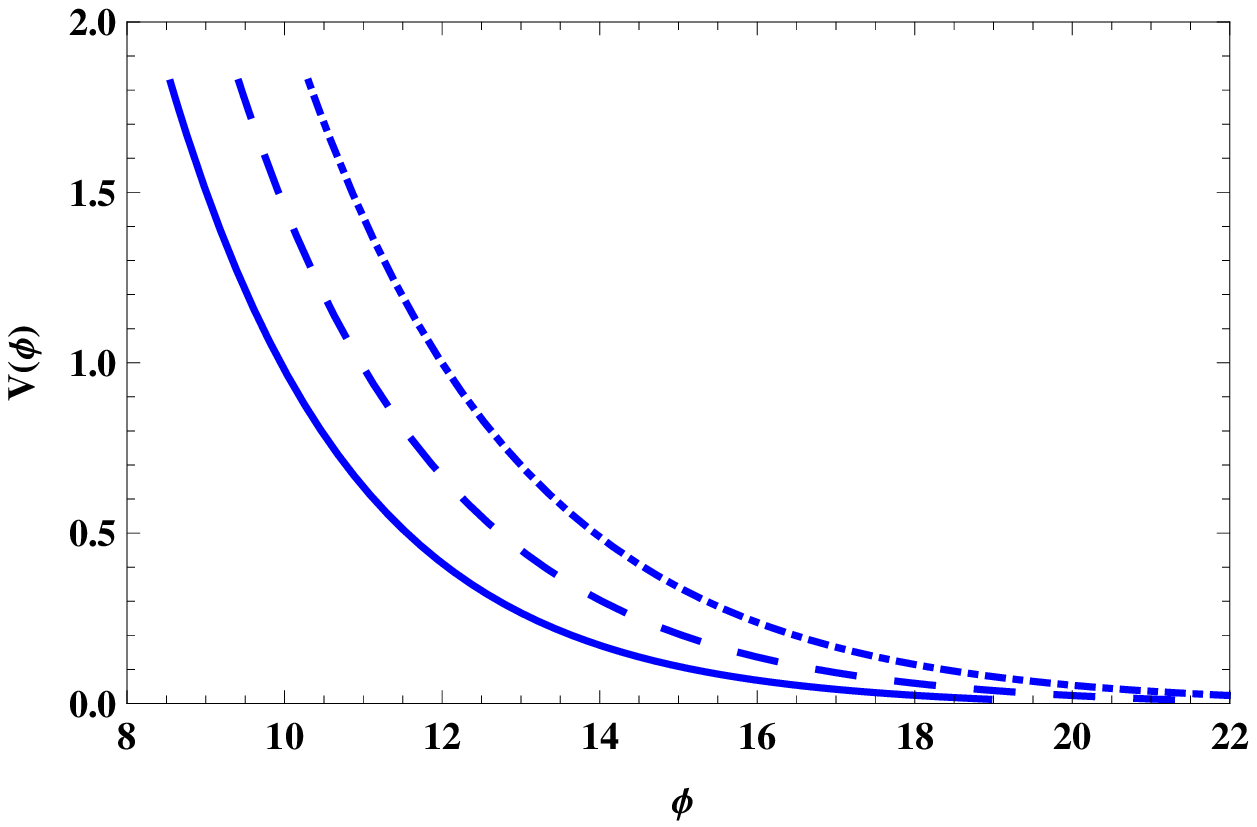}\label{SC01} }
\hspace*{5mm}
\subfigure[Potential for $\mathcal{H}_2^{2}=10^{31}$]{\includegraphics[width=5cm]{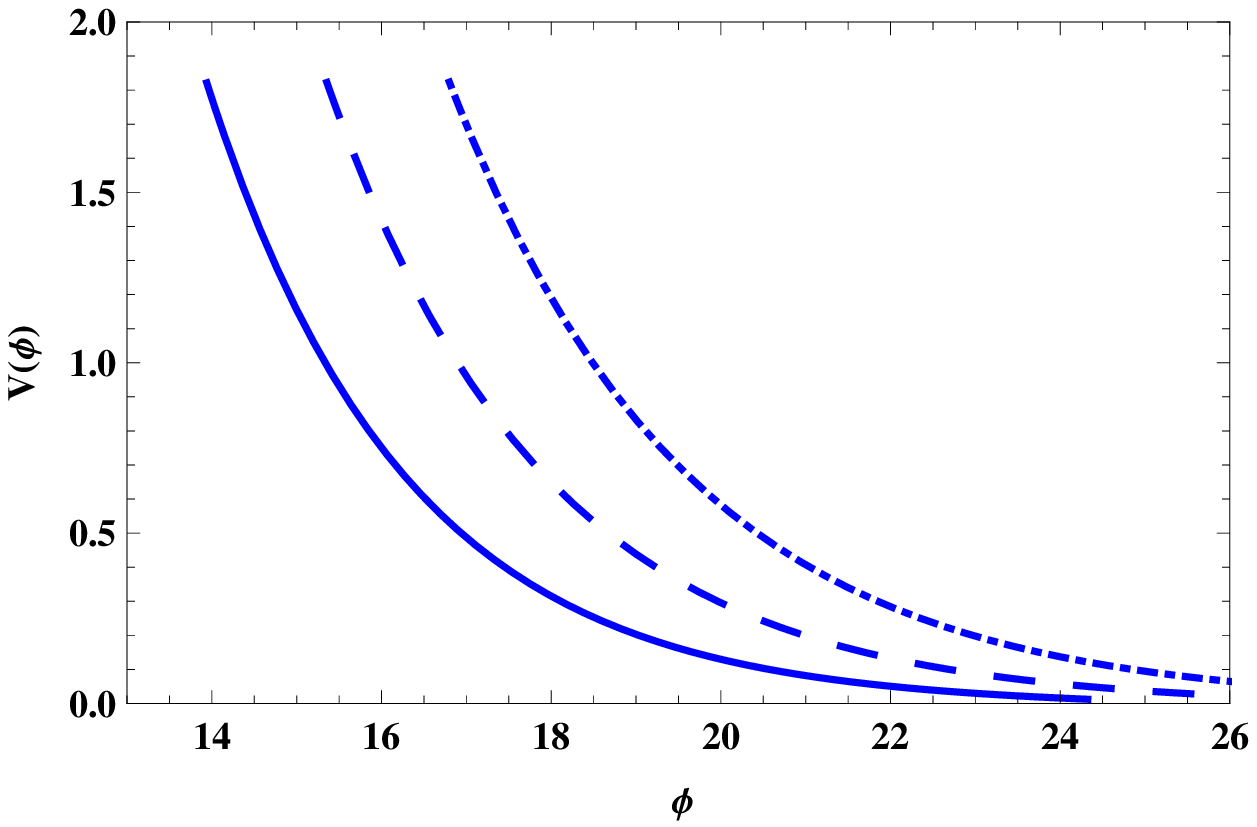}\label{SC02} }
\hspace*{5mm}
\subfigure[Potential for $\mathcal{H}_2^{2}=10^{32}$]{\includegraphics[width=5cm]{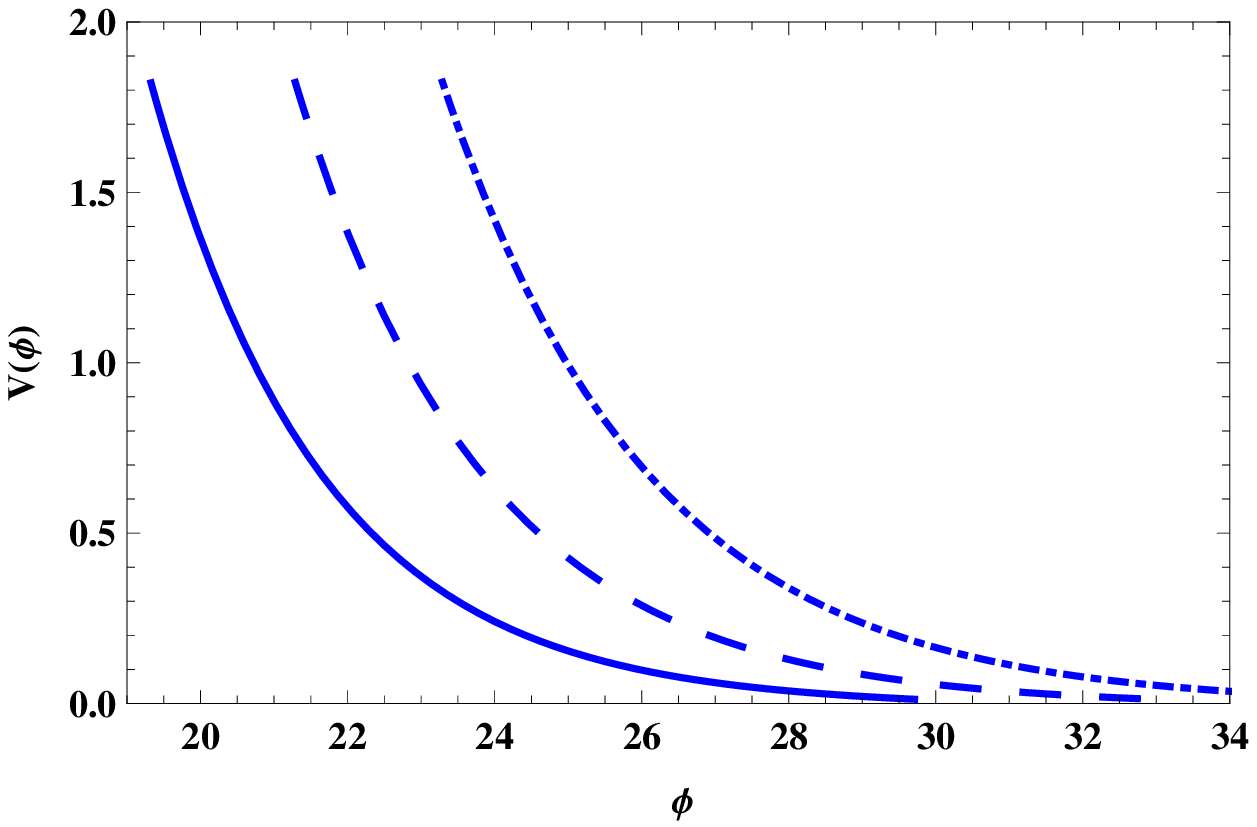}\label{SC03} }
\caption{ The potential for different values of the number of the scalar spectral index as $n_s=0.9528$ (solid line), $0.9611$ (large-lashed line), and $0.9675$ (dotted line). The vertical axis indicates the potential ($\times 10^{-64}$), and the horizon axis denotes the scalar field ($\times 10^{-14}$). }\label{F02}
\end{figure}

\end{widetext}
Recent observational data of WMAP9 + eCMB + BAO + $ H_{0} $, the Planck-2013 satellite, and Planck2013 + WP + highL + BAO, show that, the scalar spectral index is about $n_s = 0.9608 \pm 0.0080 $, $0.9635 \pm 0.0094$, and $ 0.9608 \pm 0.0054 $, respectively. Therefore, from Eq.\;(\ref{32}), one could arrive at the number of \textit{e}-folds predicted by the model and then get the tensor spectral index. The model predictions for the number of \textit{e}-folds and $n_T$ are displayed in Table \ref{Tab03} for different values of $n_s$.  By increasing the scalar spectral index, the number of \textit{e}-folds increases; however, the magnitude of the scalar spectral index reduces.

As in the previous case, the tensor perturbation for this case could be derived as well. Therefore, by using the tensor perturbation expression, one could easily determine the free parameter $s$.
Table \ref{Tab03} provides  the constraint on variable $s$ for different values of the scalar spectral index. The model supports both positive and negative values of $s$. \\
The potential for this case has been expressed in Eq.\;(\ref{28}), which could be plotted by using the results of Table \ref{Tab03}. In order to have a usual behavior for the potential, the negative values of $s$ are picked out. Finally, the potential is plotted for three different values of $\mathcal{H}_2$ in Figs. \ref{SC01}, \ref{SC02}, and \ref{SC03}. In each figure, we have plotted the evaluation of the potential for three different values $n_s$. In Fig.\ref{SC01}, it is seen that the potential has a decreasing behavior and tends to zero by increasing the scalar field.
\subsubsection{Attractive behavior}
As a final step of the first typical example, the attractor behavior of the solution is considered. From Eq.\;(\ref{16}), the perturbation $\delta H(\phi)$ is obtained as
\begin{equation}\label{35}
\delta H(\phi) = \delta H(\phi_i) \exp \left( {9\mathcal{H}_2^2 \over 4s^2} \Big[ \exp(2s\phi) - \exp(2s\phi_i) \Big] \right).
\end{equation}
By taking a negative value of parameter $s$, it is mentioned that the scalar field increases by passing time. Therefore, the exponential term on the right-hand side of Eq.(\ref{35}) reduces, which gives rise to that, the perturbation term $\delta H$ becoming smaller by passing time. Then, it seems that the second example of the model possesses the attractive behavior as well.
\section{Parameters based on BICEP2}

BICEP2 is the second generation of the background imaging cosmic extragalactic polarization (BICEP) instrument, which is
placed at the South Pole. It is designed to measure the polarization of the CMB on an angular scale of
$1$ - $5$, near the peak of the \textit{B}-mode polarization signature of primordial gravitational waves from
cosmic inflation. BICEP2 has completed three years of observation (2010-2012). The report given in
March 2014 stated that BICEP2 has detected \textit{B} modes from gravitational waves of the early Universe. An
announcement was made on March 17, 2014, that BICEP2 has detected \textit{B} modes at the level of
$r=0.20^{+0.07}_{-0.05}$ (for more details about the results and experiment, refer to Ref. \cite{3}).
In this section we are going to use the BICEP2 result for our model and find out what happens to the parameters. 
By utilizing the same process as the previous section, the parameters could be derived for both cases, in which the results 
are expressed in Table \ref{TabIII}.
\begin{widetext}

\begin{table}[ht]
  \centering
  \begin{tabular}{lp{2.5cm}p{2.5cm}p{2.5cm}p{2.5cm}}
     \toprule[1.5pt] \\ [-2mm]
         $n_s$     & $0.9528$ \qquad  & $0.9641$ \qquad  & $0.9654$ \qquad \\[0.5mm]
              \midrule[1pt] \\[-2mm]
     $\mathcal{H}_1$($N=55$) \qquad\qquad &  $1.14\times 10^{-5}$     \qquad  & $2.26\times 10^{-6}$     \qquad &  $2.88\times 10^{-7}$    \qquad \\[2mm]
     $\mathcal{H}_1$($N=60$) \qquad\qquad &  $3.70\times 10^{-4}$     \qquad  & $1.49\times 10^{-4}$     \qquad &  $4.94\times 10^{-5}$    \qquad \\[2mm]
     $\mathcal{H}_1$($N=65$) \qquad\qquad &  $2.76\times 10^{-3}$     \qquad  & $1.52\times 10^{-3}$     \qquad &  $7.49\times 10^{-4}$    \qquad \\[1.5mm]
     \midrule[1pt] \\[-2mm]
     \midrule[1pt] \\[-2mm]
     $n_s$     & $0.9528$ \qquad  & $0.9611$ \qquad  & $0.9675$ \qquad \\[0.5mm]
     \midrule[1pt] \\[-2mm]
     $s/m_p$       \qquad  \qquad  & $\pm 1.24\times 10^{-5}$ \qquad  & $\pm 1.12\times 10^{-5}$ \qquad & $\pm 1.02\times 10^{-5}$ \qquad \\[1.5mm]
     \bottomrule[1.5pt]
   \end{tabular}
  \caption{\footnotesize The parameters of the model based on the latest observational data about the scalar-to-tensor ratio, namely, BICEP2 as $r=0.20$. The other constant values are taken the same as previous cases.}\label{TabIII}
\end{table}
\end{widetext}
Note that the spectra indices remain unchanged, because they are independent of $r$, and we have different values for $\mathcal{H}_1$ in the first case and $s/m_p$ in the second case. In a comparison, one could realize that the obtained parameters by using BICEP2 results are smaller than the corresponding parameters which are obtained by using Planck-2013 data. However, the general behavior of potentials are the same as previous cases.
\section{Conclusion}

 By using the Hamilton-Jacobi approach, we have reconsidered inflationary cosmology by this assumption that the scalar field dominating the Universe, is a tachyon scalar field. After deriving the general equation of the model, it was supposed that the Hubble parameter could be defined as a function of the scalar field. By utilizing this approach, the main parameters of the model could easily be derived. The slow-roll parameters and the potential of the model were acquired as a function of the Hubble parameter. In the next step, to arrive at more specific results, two typical examples were considered, in which the Hubble parameter was introduced as a power-law and exponential function of the scalar field. By using recent observational data of the Planck-2013 satellite, WMAP9 + eCMB + BAO + $H_{0}$ and Planck2013 + WP + highL + BAO , about the scalar spectral index, and BICEP2 about the tensor-to-scalar ratio, the free parameters of the model are determined, and then the appropriate potential for each case was obtained.
 As first a case, the power-law case was considered, in which, by using the recent observational data for the scalar spectral index and the usual value of the number of \textit{e}-folds, a constraint on variable $n$ has been obtained.
The consequence has shown that, there is $n\cong -2.7, -2.2$, and $-1.9$, respectively, for $N=55, 60$, and $65$. These values of $n$ lead to tensor a spectral index about $-0.02$, which is in good agreement with observational data. Another constraint came from the amplitude of the perturbation for constant coefficient $\mathcal{H}_1$. By using these results, a schematic picture of the potential was prepared which shows that the potential decreases by increasing the scalar field.

In the second case, the Hubble parameter was described as an exponential function. The scalar spectral index relation for this case depended only on the number of \textit{e}-folds. Therefore, a constraint on $N$ was acquired by utilizing the observational data of $n_s$. It was shown that the number of \textit{e}-folds should be about 42-61. The tensor spectral index was only related to the number of \textit{e}-folds too. Therefore, the observational constraint on $N$ came to the tensor spectral index about $-0.019$, which is acceptable in comparison to observational data. The same as the previous case, by using the amplitude of the perturbation, we got another constraint for variable $s$. It was shown that this variable could be positive or negative. In order to plot the potential of this case, one needs to determine the coefficient $\mathcal{H}_2$. This parameter was not derived from observational data; however, there is another condition for $\mathcal{H}_2$, which comes from positivity of the scalar field. Then, for some proposed value of $\mathcal{H}_2$ and different values of $n_s$, the potential was depicted. It was found that the model is able to predict a kind of potential which decreases by enhancement of the scalar field.

Constraining the parameters of the model performed by using Planck-2013 and BICEP2 data, it was found out that the only changes appear in the values of $\mathcal{H}_1$, in the first case, and $s/m_p$, in the second case, and other parameters such as spectra indices remain unchanged. These parameters get a larger value by using the Planck-2013 data. It is expressed that the general behavior of the potential for both cases is the same.

The attractive behavior of the model was considered for each case. The results show a desirable situation, in which the homogeneous perturbation of the Hubble parameter $\delta H$ decreases by increasing the scalar field (or passing time). The final results demonstrate that the attractive behavior could be satisfied for both proposed cases.

\end{document}